\documentclass[journal]{IEEEtran}

\ifCLASSINFOpdf
\else
   \usepackage[dvips]{graphicx}
\fi
\usepackage{url}

\usepackage{graphicx}
\usepackage{array}
\usepackage{multirow}
\usepackage{lipsum}
\usepackage{bm}
\usepackage{amssymb}
\usepackage{amsmath}

\DeclareMathOperator*{\argmin}{arg\,min}
\usepackage{scalerel}
\usepackage{balance}

\usepackage[pagebackref=true,breaklinks=true,colorlinks,bookmarks=false]{hyperref}

\begin{document}

\title{RODIAN: Robustified Median}

\author{Seong Hun Lee and Javier Civera
\thanks{This work was supported in part by the Spanish government (project PGC2018-09637-B-I00) and in part by the Aragon regional government (DGA\_FSE T45\_20R).}
\thanks{The authors are with the University of Zaragoza, Zaragoza 50018, Spain (e-mail: seonghunlee@unizar.es; jcivera@unizar.es).}
}

% \markboth{Journal of \LaTeX\ Class Files, Vol. XX, No. X, Month 20XX}
% {Shell \MakeLowercase{\textit{et al.}}: Bare Demo of IEEEtran.cls for IEEE Journals}

\maketitle

\begin{abstract}
We propose a robust method for averaging numbers contaminated by a large proportion of outliers. 
Our method, dubbed RODIAN, is inspired by the key idea of MINPRAN \cite{minpran}:
We assume that the outliers are uniformly distributed within the range of the data and we search for the region that is least likely to contain outliers only.
The median of the data within this region is then taken as RODIAN.
Our approach can accurately estimate the true mean of data with more than 50\% outliers and runs in time $O(n\log n)$.
Unlike other robust techniques, it is completely deterministic and does not rely on a known inlier error bound.
Our extensive evaluation shows that RODIAN is much more robust than the median and the least-median-of-squares.
This result also holds in the case of non-uniform outlier distributions.
\end{abstract}


\IEEEpeerreviewmaketitle

\begin{figure*}[t]
\centerline{\includegraphics[width=0.9\textwidth]{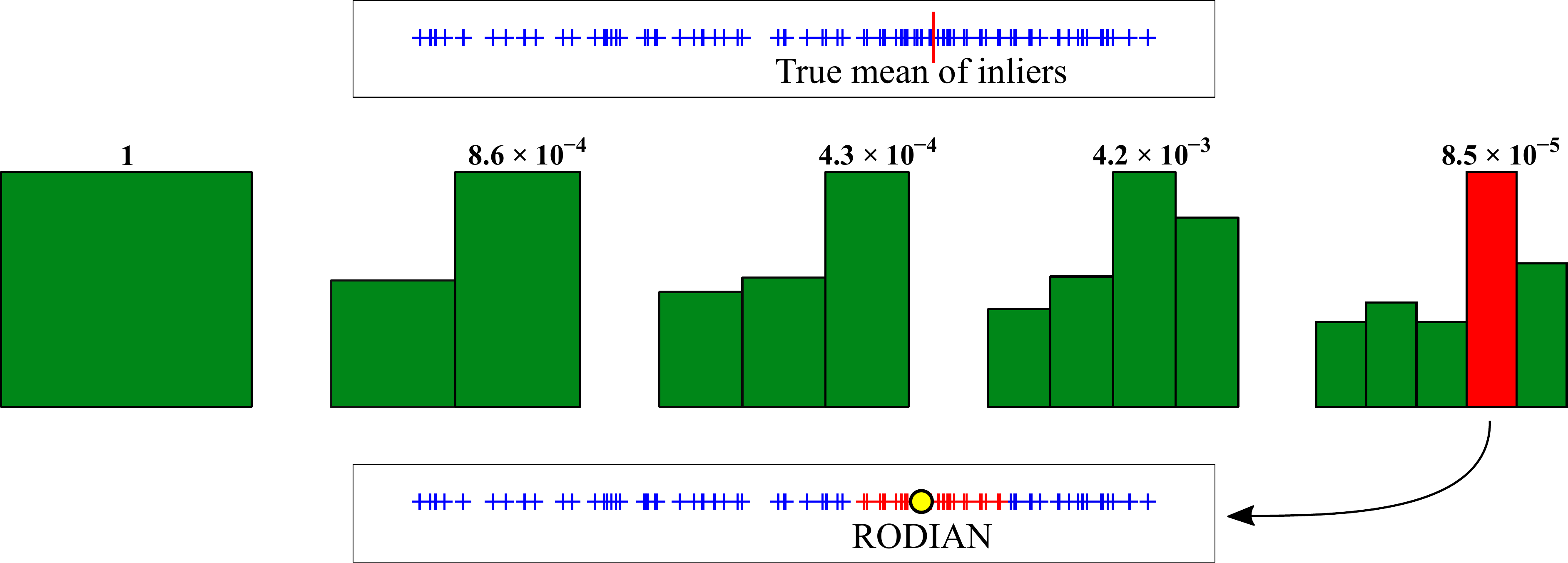}}
\caption{\textbf{Top:} 100 numbers used as input for a toy example. 
80\% of the numbers are outliers, uniformly distributed between 0 and 100.
The inliers follow $\mathcal{N}(70, 5^2)$.
\textbf{Middle:} We build multiple histograms with a varying number of bins (1--5 in this example). 
For each histogram, we find the highest bin and evaluate the probability of this occurring purely by chance, \textit{assuming that the outliers are uniformly distributed.}
These probabilities of randomness are given above each histogram.
We find the bin that produces the smallest probability (shown in red).
\textbf{Bottom:} RODIAN is the median of the numbers in this bin.
} 
\label{fig:example}
\end{figure*}

\section{Introduction}

\IEEEPARstart{A}{veraging} means finding the most representative value of a given set of data points.
For one-dimensional numbers, it can be the arithmetic mean, the median or other measures of central tendency.
All these measures have different properties, one of which is the robustness to outliers.
This is an important property to consider because outliers can severely degrade the averaging accuracy if not handled properly.
Robust averaging is a useful technique in a wide variety of domains, including pattern recognition \cite{pattern_recognition1, pattern_recognition2, pattern_recognition3}, image processing \cite{image_processing1, image_processing2, image_processing3}, 3D computer vision \cite{3d_vision1, robust_single_rotation_averaging, cui2015global}, biomedical engineering \cite{biomedical_engineering1,biomedical_engineering2,biomedical_engineering3}, economics/econometrics \cite{economics, econometrics1, econometrics2}, information science \cite{information_science1, information_science2, information_science3}, environmental studies \cite{environmental_study1, environmental_study2}, geochemistry \cite{geochemistry1, geochemistry2}, forensic science \cite{forensic_science1}, psychology \cite{psychology1} and database research \cite{database1}.

If the given data set contains a small number of outliers, it may be sufficient to use the median, as, contrary to the mean, it is robust up to a certain outlier ratio \cite{central_tendency}.
The median can be considered as a specific case of the alpha-trimmed mean with $\alpha=0.5$.
The alpha-trimmed mean \cite{alpha_trim} of $n$ numbers is defined as the arithmetic mean after truncating the largest $\frac{\alpha}{2}n$ and the smallest $\frac{\alpha}{2}n$ elements.
This method assumes that the outliers are likely to be located at the high and low ends of the sorted data.
As a result, it fails when a large number of outliers are located mostly on one side of the long tails. 

To handle such cases, a more elaborate method should be used.
One popular example is the maximum likelihood-type estimator (M-estimator) \cite{huber}.
Paired with iteratively reweighted least squares (IRLS) \cite{irls}, it can effectively downweight the influence of outliers.
However, M-estimators, such as the Huber function, often require a control parameter to be carefully tuned to the inlier error distribution.
Also, their robustness strongly depends on the initial seed, and accurate initialization in the presence of many outliers is already a non-trivial problem in and of itself.

Another popular robust estimation method is RANSAC \cite{ransac}.
It involves random sampling, but it can be made deterministic for the 1D averaging problem if we simply pick every number as a sample once.
While this method can handle a very large number of outliers, it incurs a computational cost of $O(n^2)$ and requires the prior knowledge of the inlier error bound.

The least-median-of-squares (LMedS) \cite{lmeds}, on the other hand, does not require any prior knowledge.
For the 1D averaging problem, the LMedS can be obtained by finding the data point that yields the smallest median deviation from the rest.
This would involve $O(n^2\log{n})$ computations.
Like the median, the LMedS has a breakdown point of 50\%.

Another method that does not rely on a known inlier error bound is MINPRAN \cite{minpran}.
This method is more robust than the LMedS, as it can handle more than 50\% outliers.
However, it is slower than the LMedS and has a random nature.

In this work, we propose RODIAN, a novel robust measure of central tendency.
Our method is inspired by the core idea of MINPRAN \cite{minpran}: 
We assume that the outliers follow a uniform distribution and find the median in the bounded region that is least likely to contain outliers only.
Unlike MINPRAN, however, our method is deterministic and runs in time $O(n\log n)$.\footnote{This is made possible because, unlike MINPRAN \cite{minpran}, we do not use random sampling and we fix the number of inlier bounds we evaluate (by fixing the number of histograms, as will be explained in Section \ref{sec:method}).}
Also, unlike RANSAC \cite{ransac} and Huber-like cost functions \cite{huber}, no parameter tuning is needed to account for different inlier distributions.
Our experiments show that RODIAN can handle more than 50\% outliers, outperforming the median and the LMedS \cite{lmeds} in terms of robustness.
We release our code at \url{https://seonghun-lee.github.io}.

The remainder of this paper is organized as follows: 
We detail our method in Section \ref{sec:method} and present the evaluation results in Section \ref{sec:results}.
In Section \ref{sec:limitation}, we discuss the limitation of our work.
Finally, conclusions are drawn in Section \ref{sec:conclusion}.

\section{Method}
\label{sec:method}
\subsection{Main Idea}
Suppose that we are given a set of numbers. 
Each number is either an inlier or an outlier, but we do not know which is which.
Assuming that the inliers are scattered around a certain number $\mu$, how can we estimate $\mu$ from this noisy, outlier-contaminated data?
Our approach is to find the most densely populated region in the data and take the median value in that region.
Now the question is how to determine this region.

One simple heuristic approach is to build a histogram and find the tallest bin.
Then, the edges of this bin correspond to the upper and lower bounds of the densest region.
This is a reasonable approach, but there is one problem: 
The histogram can be constructed in many different ways.
If we constrain the lower edge of the first bin to be the minimum value and the upper edge of the last bin to be the maximum value, then we can obtain multiple histograms by varying the number of bins.
So, which histogram is the right one to use?

Our answer to this question is that we choose the histogram with the bin that is least likely to have occurred by chance.
For any histogram, each of its bins have its associated probability of randomness.
For example, if all bins have the exact same height, we can deduce that the numbers are uniformly distributed, and thus random in this sense.
By the same token, if one of the bins is significantly taller than the others, then it is unlikely that it occurred due to the randomness.
In other words, this very tall bin has a low probability of randomness.

Essentially, what we propose is to build multiple histograms with the different numbers of bins, evaluate the probability of randomness associated with the tallest bin of each histogram, and choose the one that yields the smallest probability of randomness.
This is because the minimum probability of randomness implies the maximum probability of containing mostly inliers.
This process is illustrated in Fig. \ref{fig:example}.

The remaining question is how exactly we compute this probability of randomness.
Basically, we adopt a similar idea proposed by Stewart \cite{minpran} and compute the probability in a binomial distribution, assuming that the random outliers are uniformly distributed across the entire range.
Note that if a trial has a probability of success $p$, the probability of obtaining $k$ successes from $n$ trials is given by
\begin{equation}
    P(k)=\begin{pmatrix}n \\ k\end{pmatrix} p^k(1-p)^{n-k}.
\end{equation}
In our problem, $p$ is the probability of a random outlier falling inside the tallest bin, $k$ is the frequency of this bin, and $n$ is the size of the data. 
Since an outlier can fall inside any other bins with an equal probability, we have $p=1/b$ where $b$ is the number of bins.
Therefore, the probability of randomness associated with the bin containing $k$ numbers is given by
\begin{equation}
\label{eq:probability_of_randomness}
    P(k)=\begin{pmatrix}n \\ k\end{pmatrix} \left(\frac{1}{b}\right)^k\left(1-\frac{1}{b}\right)^{n-k}.
\end{equation}
In summary, we vary $b$, find $k$ by building a histogram, compute $P(k)$, and repeat this process until we find the value of $b$ that leads to the smallest $P(k)$.
In the next section, we discuss several strategies we came up with to improve the efficiency of the algorithm.

We note that, in contrast to our approach, MINPRAN \cite{minpran} uses the probability that \textit{at least} $k$ data points fall within the inlier region, which is technically different from \eqref{eq:probability_of_randomness}.
Empirically, we found that using this probability instead of \eqref{eq:probability_of_randomness} leads to similar results, but at a slower speed.

\subsection{Implementation Details}
\textbf{1. How many histograms do we consider?:}\\
According to the pigeonhole principle, if we set the number of bins to $n-1$, at least one bin will contain more than one number.
Therefore, one could find the theoretically optimal number of bins, $b^*$, by varying $b$ from $1$ to $n-1$, searching for the minimum $P(k)$ in Eq. \eqref{eq:probability_of_randomness}.
However, in our experiment described in Section \ref{sec:results}, we found that setting $b>20$ hardly makes any difference in the final accuracy.
We also empirically found that there is no need to try all integers between 1 and 20, as similar results could be obtained faster with $b\in\{2,3,4,5,7,9,11,14,17,20\}$.

\textbf{2. How to accelerate the histogram building process:}\\
Building multiple histograms one by one can take a long time.
For efficiency, we precompute a table that matches the bin edges and the bin indices of all the histograms. 
This process is explained Fig. \ref{fig:bin_edges}.
In order to reuse this table on any data, it must be agnostic of the input.
To this end, we normalize the input data such that its range becomes $[0,1]$.
This way, all edges get predetermined values between $0$ and $1$.

\break

\subsection{Summary}
\begin{figure}[t]
\centerline{\includegraphics[width=0.65\columnwidth]{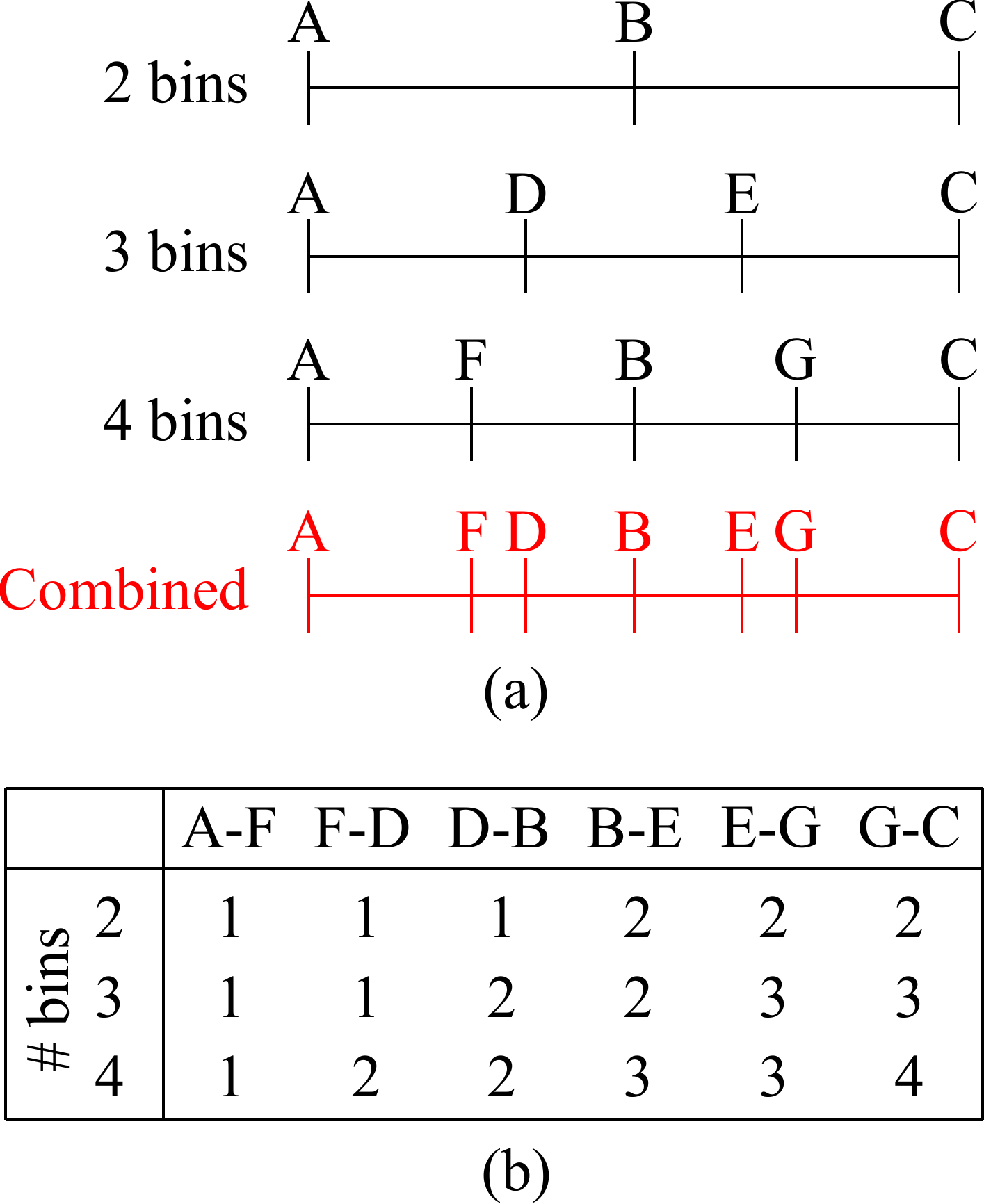}}
\caption{
This example illustrates the process of precomputing the tabular data for building histograms:
\textbf{(a)} Suppose that we want to build three histograms with 2, 3 and 4 bins.
We collect the bin edges of all histograms (without duplicates) and sort them in a single array, \textit{i.e.}, $[A,F,D,B,E,G,C]$.
\textbf{(b)} We assign the corresponding bin index to each region bounded by two successive edges in the combined array. 
For example, a number between edge $F$ and $D$ would fall inside the $1^\text{st}$ bin in the 2-bin and 3-bin histogram, and $2^\text{nd}$ bin in the 4-bin histogram.
These indices are given in the F-D column.
By checking which region a number falls in, we can immediately find the corresponding bin index in each histogram. 
} 
\label{fig:bin_edges}
\end{figure}

\begin{enumerate}\itemsep0em
    \item Precompute a table, as described in Fig. \ref{fig:bin_edges}, with $A=0$, $B=0.5$, $C=1$, etc.
    For the number of bins, we use $b\in\{2,3,4,5,7,9,11,14,17,20\}$.
    \item Sort and normalize the input such that its range is between $0$ and $1$, \textit{i.e.,}
    \begin{equation}
        x_i \gets \frac{x_i - x_\text{min}}{x_\text{max}-x_\text{min}} \quad \text{for} \ i=1, 2, \cdots, n.
    \end{equation}
    \item For each data point, use the precomputed table to find the corresponding bin index in each histogram.
    \item For each histogram, find the frequency of the tallest bin and the associated probability of randomness (Eq. \eqref{eq:probability_of_randomness}).
    \item Find the histogram that leads to the smallest probability. 
    \item In that histogram, find the median of the numbers that fall inside the tallest bin.
    \item Unnormalize this median. The final value is RODIAN.
\end{enumerate}
In Step 4, we discard a histogram if multiple bins have the same maximum frequency.
If, by any chance, all histograms are discarded, we simply take the median of the original input.

\textbf{Time analysis:} The time complexity of Step 2 and 6 is $O(n \log n)$.
The other steps run in either $O(1)$ or $O(n)$.
Hence, the total time complexity is $O(n \log n)$.

\section{Results}
\label{sec:results}
\begin{figure*}[t]
\centerline{\includegraphics[width=\textwidth]{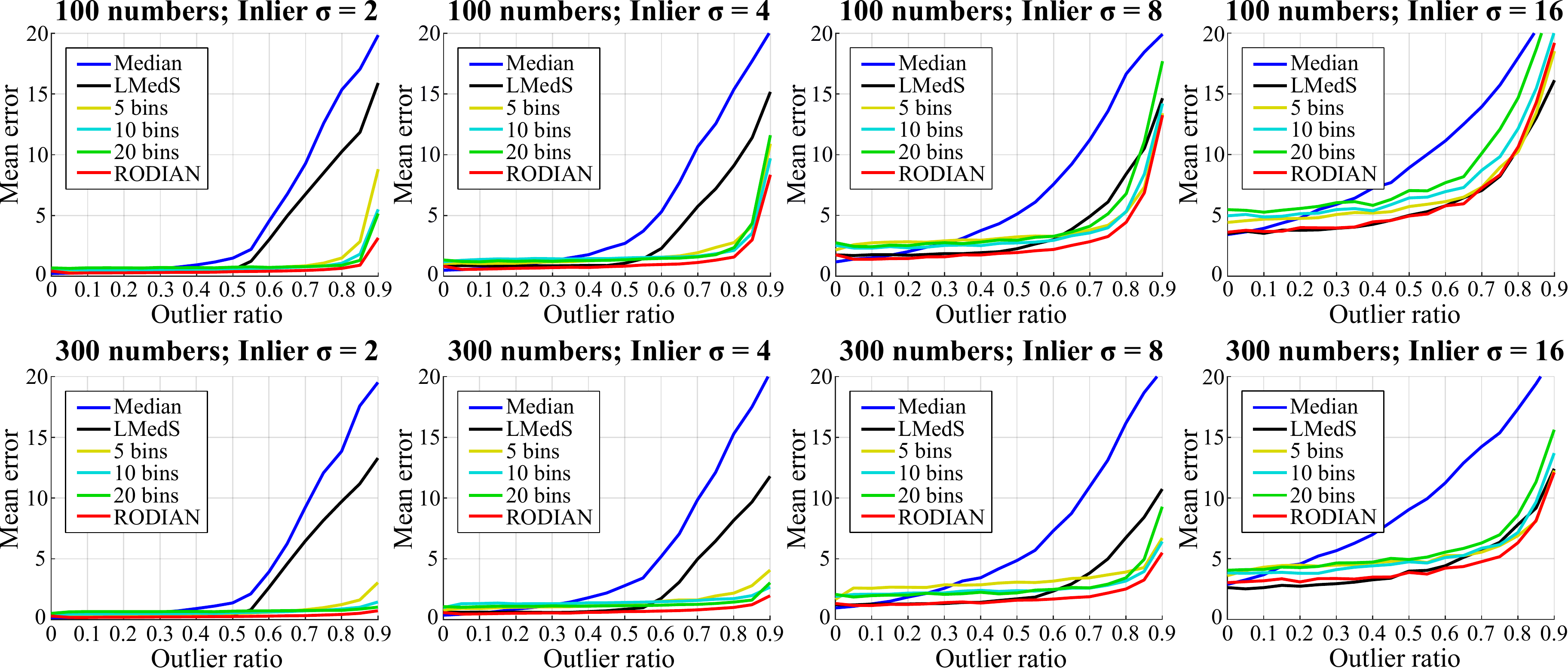}}
\caption{Mean error comparison under a \textbf{uniform outlier distribution}: 
The evaluation is carried out with 100 numbers (top row) and 300 numbers (bottom row).
The numbers are generated such that the inliers follow $\mathcal{N}(\mu, \sigma^2)$ where $0<\mu<100$ and $\sigma\in\{2, 4, 8, 16\}$, and the outliers follow $\mathcal{U}(0, 100)$.
If any number is outside a range $[0, 100]$, it is removed and regenerated until all numbers are within this range.
Each data point in the graph represents the mean error of 1000 independent runs.
It shows that, across different inlier noise levels, RODIAN is the most robust to outliers.
When the inlier noise is small, the breakdown point of RODIAN is well over 80\% (see the first column). 
Also, RODIAN is generally more accurate than the fixed-histogram approach.
}
\label{fig:uniform_outliers}
\end{figure*}

\begin{figure*}[t]
\centerline{\includegraphics[width=\textwidth]{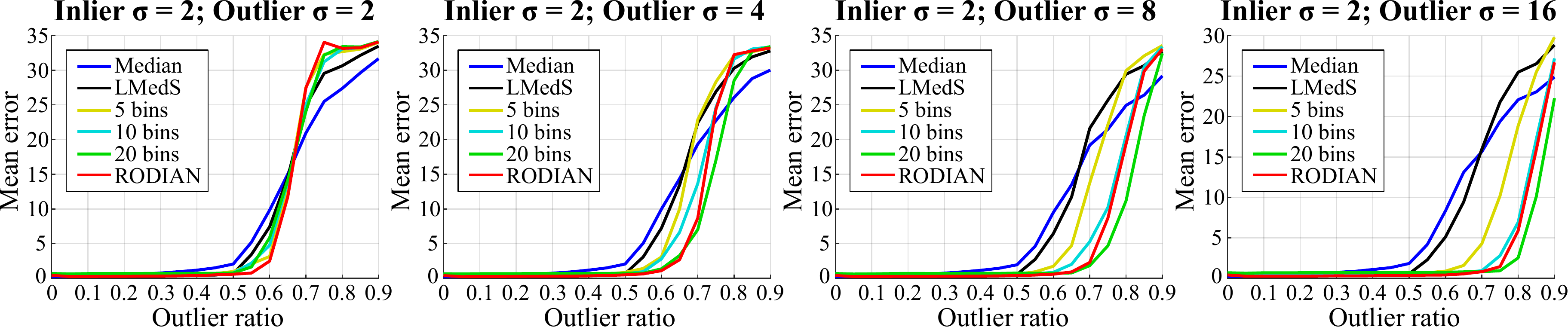}}
\caption{Mean error comparison under a \textbf{uniform + Gaussian outlier distribution}:
The evaluation is carried out with 100 numbers.
The numbers are generated using the same procedure described in Fig. \ref{fig:uniform_outliers}, except that half of the outliers now follow $\mathcal{N}(\mu_\mathrm{outlier}, \sigma_\mathrm{outlier}^2)$, $0<\mu_\mathrm{outlier}<100$.
We observe that RODIAN has a higher breakdown point than the median and the LMedS.
When the outlier ratio is very high, the fixed-histogram approach with 20 bins produces smaller errors than RODIAN.
However, they both are well beyond the breakdown point there and comparisons are meaningless, as errors are driven by outliers.
} 
\label{fig:gaussian_outliers}
\vspace{-1em}
\end{figure*}

We compare RODIAN with three other methods:
\begin{enumerate}
    \item Median,
    \item Least-median-of-squares (LMedS) \cite{lmeds}, estimated as the data point with the smallest median (squared) distance to the rest, \textit{i.e.},
    
    \begin{equation}
        \mathrm{LMedS} = \underset{\scaleto{x_i}{5pt}}{\argmin} \ \underset{j}{\mathrm{med}} \left(x_i - x_j\right)^2,
    \end{equation}
    
    \item Median of the tallest bin of a fixed histogram,  obtained in the following way: (i) Build a histogram with a fixed number of bins, (ii) Collect all the numbers that fall inside the tallest bin, (iii) Compute their median.
\end{enumerate}
We present the results on synthetic data with two different outlier distributions: a uniform distribution (Fig. \ref{fig:uniform_outliers}) and a mixture of a uniform and a Gaussian distribution (Fig. \ref{fig:gaussian_outliers}).
In both cases, RODIAN outperforms the rest in terms of robustness.
Especially, Fig. \ref{fig:gaussian_outliers} shows that even though we assumed a uniform outlier distribution in our derivation of RODIAN, it can still handle non-uniform outliers relatively well if $\sigma_\mathrm{outlier}$ is larger than $\sigma_\mathrm{inlier}$.

\begin{figure}
\centerline{\includegraphics[width=0.85\columnwidth]{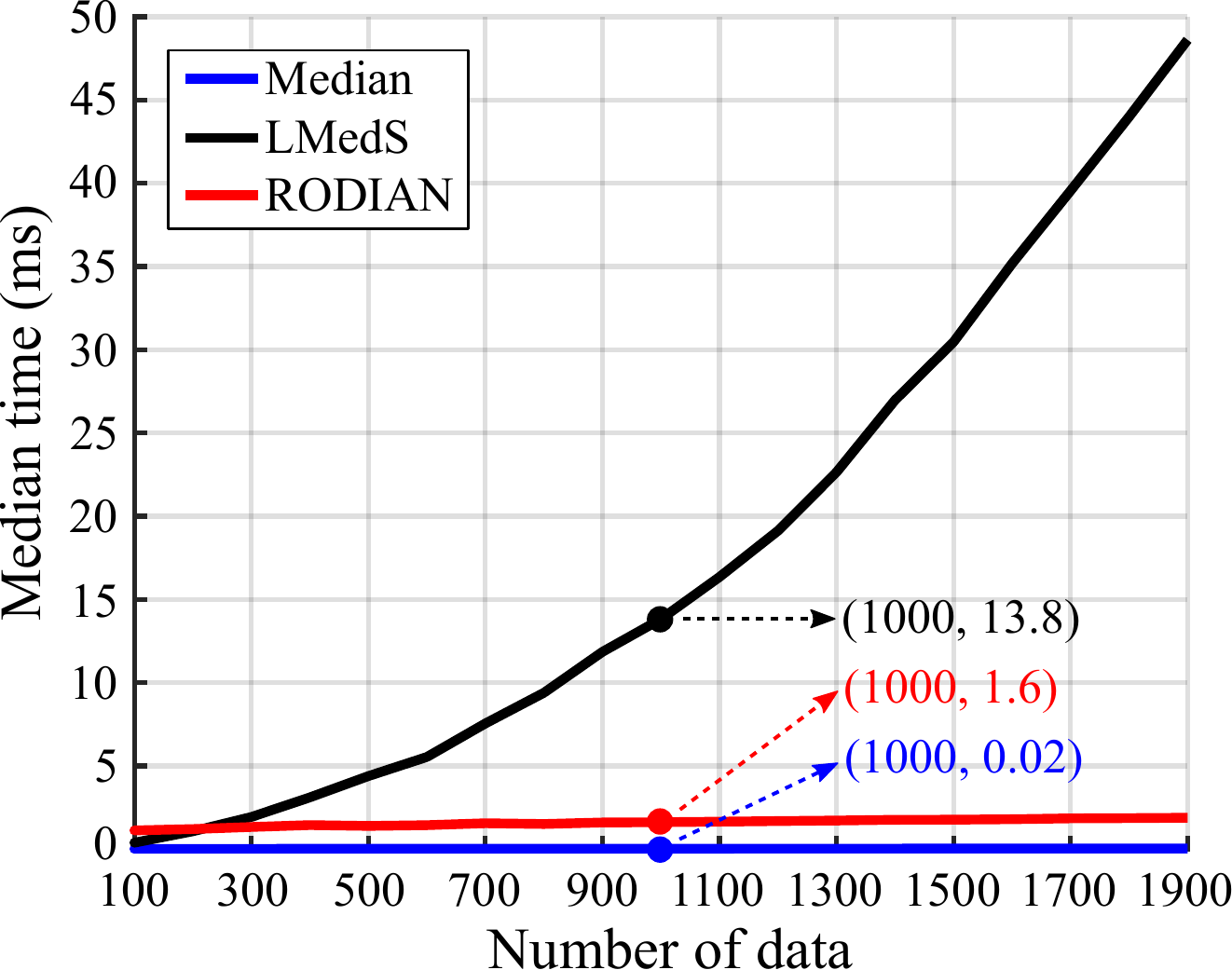}}
\caption{Median computation times (ms) of 1000 runs: The time complexity of the median, LMedS and RODIAN is $O(n\log n)$, $O(n^2\log n)$ and $O(n\log n)$, respectively.
The median is always the fastest, and RODIAN is faster than the LMedS when \#data $\gtrapprox$ 200.
All methods are implemented in MATLAB and run on a laptop with an Intel's 4th Gen i7 CPU (2.8 GHz).}
\label{fig:time}
\end{figure}

Table \ref{myTab} compares the accuracy of RODIAN and the fixed-histogram approach for low to moderate outlier ratios.
It again demonstrates the advantage of using RODIAN over a fixed histogram.
In Fig. \ref{fig:time}, we plot the mean computation times of the median, LMedS and RODIAN.
We observe that RODIAN is much more scalable than the LMedS.

\section{Limitation}
\label{sec:limitation}
The main limitation of RODIAN is that its accuracy slightly drops when there are too few outliers (see Table \ref{myTab}). 
This happens because the densest region of the inlier distribution is not always well aligned with the location of the true mean.
While this is certainly not a favorable property, the average error increase is relatively small (around 10\% of the standard deviation of the inliers in Table \ref{myTab}).
We believe that this is a tolerable level in outlier-prone situations, which is the main domain we target in this work.

One potential solution is to detect when the data is outlier-free and switch to a traditional median.
If the type of the inlier distribution is known (\textit{e.g.,} Gaussian), one can use a statistical test to check if the whole data follow the inlier distribution (\textit{e.g.,} normality test \cite{thode2002testing, shapiro1965analysis}).
In this work, however, we aim to make our method generalizable to any types of inlier distribution as long as it is unimodal.
Discerning outlier-free data in such a general scenario is left for future work.

\section{Conclusion}
\label{sec:conclusion}
In this work, we presented RODIAN, a novel method for averaging outlier-contaminated numbers.
It consists of two main steps: (1) determine a bounded region in the range that would contain mostly inliers, and (2) find the median within that region.
The key idea of the first step is to assume a uniform outlier distribution and search for the region that is least likely to have occurred due to outliers.
Unlike MINPRAN \cite{minpran}, where a similar idea was used, our method is deterministic and runs in time $O(n\log n)$. 
Unlike RANSAC \cite{ransac} and Huber-like loss functions \cite{huber}, we do not need to tune a control parameter to adapt to different inlier error distributions.
Finally, unlike the median and the LMedS \cite{lmeds}, RODIAN can handle more than 50\% outliers.
An extensive evaluation demonstrates its excellent robustness, versatility and scalability.


\break

{\renewcommand{\arraystretch}{1.0}%
\begin{table}[t]
\caption{USING FIXED HISTOGRAMS vs. RODIAN}
\label{myTab}
\vspace{-1em}
\small
\setlength{\tabcolsep}{3.7pt}
\begin{center}
\begin{tabular}{|c|c|cccccc|}
\hline
\multicolumn{2}{|c|}{Outliers} &5 bins & 10 bins & 20 bins & 30 bins & 50 bins & RODIAN\\
\hline
\parbox[t]{7pt}{\multirow{6}{*}{\rotatebox[origin=c]{90}{Uniform$^a$}}} &0\% & 0.54 & 0.64 & 0.69 & 0.74 & 0.80 & \textbf{0.42} \\
& 10\% & 0.36 & 0.50 & 0.69 & 0.61 & 0.61 & \textbf{0.26} \\
& 20\% & 0.39 & 0.51 & 0.69 & 0.63 & 0.62 & \textbf{0.28} \\
& 30\% & 0.43 & 0.54 & 0.70 & 0.64 & 0.63 & \textbf{0.30} \\
& 40\% & 0.48 & 0.57 & 0.71 & 0.67 & 0.66 & \textbf{0.32} \\
& 50\% & 0.56 & 0.61 & 0.72 & 0.68 & 0.68 & \textbf{0.36} \\
\hline
\parbox[t]{8pt}{\multirow{6}{*}{\rotatebox[origin=c]{90}{Gaussian$^b$}}} &0\% & 0.54 & 0.63 & 0.69 & 0.73 & 0.79 & \textbf{0.42} \\
& 10\% & 0.46 & 0.64 & 0.63 & 0.64 & 0.68 & \textbf{0.28} \\
& 20\% & 0.47 & 0.64 & 0.65 & 0.66 & 0.69 & \textbf{0.30} \\
& 30\% & 0.47 & 0.65 & 0.68 & 0.67 & 0.72 & \textbf{0.33} \\
& 40\% & 0.50 & 0.71 & 0.77 & 0.77 & 0.85 & \textbf{0.38} \\
& 50\% & 4.36 & 2.33 & 2.29 & 1.93 & 2.49 & \textbf{1.34} \\
\hline
\multicolumn{8}{p{27em}}{$^{\phantom{A^A}}$\hspace{-12pt}
We generate 100 numbers within a range $[0, 100]$ and average them using either a fixed histogram or RODIAN.
This is repeated 10000 times.
RODIAN produces the smallest mean error.}\\
\multicolumn{8}{p{27em}}{$^a$Inliers follow $\mathcal{N}(\mu, 2^2)$ with $0<\mu<100$.}\\
\multicolumn{8}{p{27em}}{$^b$Inliers and outliers follow $\mathcal{N}(\mu_1, 2^2)$ and $\mathcal{N}(\mu_2, 4^2)$ with $0<\mu_1, \mu_2<100$.
(Note: This dataset is different from that of Fig. \ref{fig:gaussian_outliers}.)}\\
\end{tabular}
\end{center}
\end{table}}

\clearpage
\newpage
{
\balance
%\bibliographystyle{IEEEtran}
%\bibliography{IEEEabrv,mybib}


}

\end{document}